\documentclass[aps,prl,twocolumn,amssymb,showpacs]{revtex4}
\usepackage{amsmath}

\begin{document}

%%%%%%%%%%%%%%%%%%%%%%%%%%%%%%%%%%%%%%%%%%%%%%%%%%%%%%%%%%%%%%%%%%%%%
%%%%%%%%%%%%%%%%%%%%%         Title       %%%%%%%%%%%%%%%%%%%%%%%%%%%
%%%%%%%%%%%%%%%%%%%%%%%%%%%%%%%%%%%%%%%%%%%%%%%%%%%%%%%%%%%%%%%%%%%%%

\title{The Lorentz-Dirac force from QED for linear acceleration}

%%%%%%%%%%%%%%%%%%%%%%%%%%%%%%%%%%%%%%%%%%%%%%%%%%%%%%%%%%%%%%%%%%%%%
%%%%%%%%%%%%%%%%%%%%     Authors & Addresses  %%%%%%%%%%%%%%%%%%%%%%%
%%%%%%%%%%%%%%%%%%%%%%%%%%%%%%%%%%%%%%%%%%%%%%%%%%%%%%%%%%%%%%%%%%%%%
\author{ Atsushi Higuchi$^1$ and Giles~D.~R.~Martin$^2$}

\affiliation{Department of Mathematics, University of York,
Heslington, York YO10 5DD, United Kingdom\\ email: $^1$ah28@york.ac.uk,
$^2$gdrm100@york.ac.uk}

\date{July 20, 2004}

\begin{abstract}
We investigate the motion of a wave packet of a charged scalar
particle linearly accelerated by a static potential in quantum
electrodynamics. We calculate the expectation value of the
position of the charged particle after the acceleration to first
order in the fine structure constant in the $\hbar \to 0$ limit.
We find that the change in the expectation value of the position
(the position shift) due to radiation reaction agrees exactly with
the result obtained using the Lorentz-Dirac force in classical
electrodynamics. We also point out that the one-loop correction to
the potential may contribute to the position change in this limit.
\end{abstract}


\maketitle

A charged particle radiates when it is accelerated.  The resulting
change in its energy and momentum is described by the
Lorentz-Dirac (or Abraham-Lorentz-Dirac) force in classical
electrodynamics~\cite{Abraham,Lorentz,Dirac}. (See, e.g.,
Ref.~\cite{Poisson} for a modern review.) Thus, if a charge $e$
with mass $m$ is accelerated by an external 4-force $F^\mu_{\rm
ext}$, then its coordinates $x^\mu(\tau)$ at the proper time
$\tau$ obey the following equation:
\begin{equation}
m\frac{d^2x^\mu}{d\tau^2} = F^\mu_{\rm ext} + F^\mu_{\rm LD}\,,
\label{Lor}
\end{equation}
where the Lorentz-Dirac 4-force $F^\mu_{\rm LD}$ is given by
\begin{equation}
F^\mu_{\rm LD} \equiv \frac{2\alpha_c}{3}\left[
\frac{d^3x^\mu}{d\tau^3} + \frac{dx^\mu}{d\tau}\left(\frac{d^2
x^\nu}{d\tau^2} \frac{d^2x_\nu}{d\tau^2}\right)\right]\,.
\end{equation}
We have let $c=1$ and defined $\alpha_c\equiv e^2/4\pi$. Our metric
is $g_{\mu\nu} = {\rm diag}\,(+1,-1,-1,-1)$.

Although there are many ways to derive Eq.~(\ref{Lor}) in
classical electrodynamics for a point charge (see, e.g.,
Ref.~\cite{Teit}), a natural question one can ask is whether or
not this equation can arise in the $\hbar \to 0$ limit in QED. It
was found in Ref.~\cite{Higuchi2} (after an initial claim to the
contrary) that the position of a linearly accelerated charged
particle in the Lorentz-Dirac theory is reproduced by the $\hbar
\to 0$ limit of the one-photon emission process in QED in the
non-relativistic approximation. (See, e.g., Refs.~\cite{MS,beilok}
for other approaches to arrive at the Lorentz-Dirac theory from
QED.) In this Letter we describe the generalization of this work 
to a fully relativistic charged particle.
The details will be published elsewhere~\cite{HM2}.

Consider a charged particle with charge $e$ and mass $m$ moving in
the positive $z$-direction under a potential energy $V(z)$. We
assume that $V(z) = V_0 = {\rm const}.$ for $z < - Z_1$ and
$V(z)=0$ for $- Z_2 < z$ for some $Z_1$ and $Z_2$, both positive
constants.  Thus, there is non-zero acceleration only in the interval
$(-Z_1,-Z_2)$.  The external 4-force in Eq.~(\ref{Lor}) generated
by this potential is given by $F^t_{\rm ext} = - V'(z)\,dz/d\tau$
and $F^z_{\rm ext} = - V'(z)\,dt/d\tau$ with $F^x_{\rm ext} =
F^y_{\rm ext} = 0$.  The Lorentz-Dirac 4-force can be given by
$F^t_{\rm LD} = F_{\rm LD}\,dz/d\tau$, $F^z_{\rm LD} = F_{\rm
LD}\,dt/d\tau$, $F^x_{\rm LD} = F^y_{\rm LD} = 0$ with
\begin{equation}
F_{\rm LD}  \equiv \frac{2\alpha_c}{3}\gamma \frac{d\ }{dt}
(\gamma^3\ddot{z})\,.  \label{LDLD}
\end{equation}
A dot indicates the derivative with respect to $t$. We have
defined $\gamma \equiv (1-\dot{z}^2)^{-1/2}$ as usual.

Suppose that this particle would be at $z=0$ at time $t=0$ {\em if
the Lorentz-Dirac force was absent (i.e. if $e=0$)}.  
The true position at $t=0$,
which we denote $\delta z$ and call the position shift, can
readily be found to lowest non-trivial order in $F_{\rm LD}$ by
treating the Lorentz-Dirac force as perturbation. (It was proposed
in Ref.~\cite{Landau} that one should treat the Lorentz-Dirac
force as perturbation.) The
calculation can be facilitated by using the fact that the total
energy, $m\,dt/d\tau + V(z)$, changes by the amount of work done
by the Lorentz-Dirac force.  Thus,
\begin{eqnarray}\label{0}
\frac{d}{d\dot{z}}\frac{m}{\sqrt{1-\dot{z}^2}} \delta \dot{z}
 + \frac{dV(z)}{dz}\delta z
 &=& m\gamma^3\dot{z}^2 \frac{d}{dt}\left[\frac{\delta
 z}{\dot{z}}\right]\nonumber \\
& = & \int_{-\infty}^{t} F_{\rm LD}(t')\dot{z}(t') dt'.
\end{eqnarray}
Rearranging and integrating, and then changing the order of
integration, we obtain
\begin{equation}
\delta z = -\frac{\dot{z}|_0}{m}\int^0_{-\infty}
\left(\int^{t}_{0}\frac{1}{\gamma^3(t')[\dot{z}(t')]^2}dt'
\right)F_{\rm LD}\frac{dz}{dt}dt\,,\label{deltaz}
\end{equation}
where the final velocity is denoted $\dot{z}|_0$. We will show
that this position shift is reproduced to order $e^2$
in QED in the $\hbar \to 0$ limit.

The Lagrangian density for the corresponding field-theoretic model
is
\begin{equation}
{\cal L} = (D_\mu\varphi)^\dagger D^\mu \varphi -
(m/\hbar)^2\varphi^\dagger \varphi -
\frac{1}{4}F_{\mu\nu}F^{\mu\nu}\,,
\end{equation}
where $F_{\mu\nu} \equiv \partial_\mu A_\nu - \partial_\nu A_\mu$
and $D_\mu \equiv \partial_\mu + i(eA_\mu + V_\mu)/\hbar$. The
field $\varphi$ describes a charged scalar particle with mass $m$ and
charge $e$, the field $A_\mu$ is the electromagnetic field, and
the function $V_\mu = V(z)\delta_{\mu 0}$ is the external
potential which accelerates the charged scalar particle.

Now, let $A^\dagger({\bf p})$ be the creation operator for the
charged scalar particle with momentum ${\bf p}$ with 
$p\equiv p_z > 0$ 
in the positive $z$ region such that the corresponding
mode function $\Phi_{\bf p}(t,{\bf x})$ is well approximated by
the WKB approximation:
\begin{eqnarray}
\Phi_{\bf p}(t,{\bf x}) & \approx &
\sqrt{\frac{p}{\kappa_p(z)}}\exp \left[ \frac{i}{\hbar}\int_0^z
\kappa_p(\zeta)\,d\zeta\right] \nonumber \\ && \times \exp \left[
\frac{i}{\hbar}{\bf p}_\perp\cdot{\bf x}_\perp -
\frac{i}{\hbar}p_0 t\right]\,, \label{WKB}
\end{eqnarray}
 where ${\bf
p}_\perp \equiv (p_x,p_y)$, $p_0 \equiv (p^2+{\bf p}_\perp^2 +
m^2)^{1/2}$, ${\bf x}_\perp \equiv (x,y)$ and where
$\kappa_p(z)\equiv\{[p_0-V(z)]^2 -m^2-{\bf p}_\perp^2\}^{1/2}$ is
the $z$-component of the momentum of the classical particle at
$z$ with $e=0$. If the initial state in the interaction picture is
$A^\dagger({\bf p})|0\rangle$, then, to order $e^2$, it evolves as
follows:
\begin{eqnarray}
A^\dagger({\bf p})|0\rangle & \to & [1 + i{\cal F}({\bf
p})]A^\dagger({\bf p})|0\rangle\nonumber \\ &&  + \frac{i}{\hbar}
\int \frac{d^3{\bf k}}{(2\pi)^3 2k_0} {\cal A}^\mu({\bf p},{\bf
k}) a_\mu^\dagger({\bf k})A^\dagger({\bf P})|0\rangle\,, \nonumber
\\ \label{trans}
\end{eqnarray}
where the creation operators $a^\dagger_\mu({\bf k})$ for the
photons with momenta $\hbar {\bf k}$ in the Feynman gauge satisfy
\begin{equation}
\left[ a_\mu({\bf k}),a^\dagger_\nu({\bf k}')\right] =
-g_{\mu\nu}(2\pi)^32\hbar k_0\delta({\bf k}-{\bf k}')\,.
\end{equation}
Note that the momentum $\hbar {\bf k}$ of the photon is of order
$\hbar$ because the wave number ${\bf k}$ rather than the momentum
has the classical limit. The ${\cal F}({\bf p})$ is the
forward-scattering amplitude coming from the one-loop diagram,
which will not be evaluated explicitly. The momentum ${\bf P}$ of
the charged particle in the final state is determined using
conservation of the transverse momentum, ${\bf p}_\perp = {\bf
P}_\perp + \hbar {\bf k}_\perp$, and energy conservation, $p_0 =
P_0 + \hbar k_0$. The expectation value of the $z$-coordinate of the
wave packet at $t=0$ is given in terms of the forward scattering
amplitude ${\cal F}({\bf p})$ and the photon emission amplitude
${\cal A}_\mu({\bf p},{\bf k})$ as we describe next.

Let the initial state $|i\rangle$ be given by
\begin{equation}
| i\rangle = \int \frac{d^3{\bf p}}{\sqrt{2p_0}(2\pi\hbar)^3}
f({\bf p})A^\dagger({\bf p})|0\rangle\,,
\end{equation}
where the function $f({\bf p})$ is sharply peaked about a
momentum in the positive $z$-direction with width of
order $\hbar$. The operators $A^\dagger({\bf p})$ are normalized
so that the condition $\langle i\,|\,i\rangle = 1$ leads to
\begin{equation}
\int\frac{d^3{\bf p}}{(2\pi\hbar)^3}|f({\bf p})|^2 = 1\,.
\end{equation}
Hence, the function $f({\bf p})$ can be regarded as the
one-particle wave function in the momentum representation.  This
suggests that the position at $t=0$ {\em in the absence of
radiation} is given by
\begin{equation}
\langle z\rangle_0 = \frac{i\hbar}{2}\int \frac{d^3{\bf
p}}{(2\pi\hbar)^3} f^*({\bf p})
\stackrel{\leftrightarrow}{\partial}_p f({\bf p})\,,
\end{equation}
where $\stackrel{\leftrightarrow}{\partial}_p
 \equiv \stackrel{\rightarrow}{\partial}_p - \stackrel{\leftarrow}
{\partial}_p$. (See Appendix B of Ref.\ \cite{Higuchi2} for a
derivation.) We assume in this Letter that $f({\bf p})$ is real
for simplicity.  Hence, we have $\langle z\rangle_0 = 0$
as required.

The final state resulting from the initial state $|i\rangle$ can
be found from Eq.~(\ref{trans}) as
\begin{eqnarray}
| f\rangle & = & \int\frac{d^3{\bf p}}{\sqrt{2p_0}(2\pi\hbar)^3}
F({\bf p}) A^\dagger({\bf p})|0\rangle \nonumber
\\
&& + \frac{i}{\hbar}\int \frac{d^3{\bf k}}{2k_0(2\pi)^3}
\int\frac{d^3{\bf P}}{\sqrt{2P_0}(2\pi\hbar)^3}\nonumber \\ && \ \
\ \ \ \ \ \ \ \times G^\mu({\bf p},{\bf k}) a_\mu^\dagger({\bf
k})A^\dagger({\bf P})|0\rangle\,,
\end{eqnarray}
where we have defined $F({\bf p}) \equiv (1 + i{\cal F}({\bf
p}))f({\bf p})$ and $G^\mu({\bf p},{\bf k}) \equiv {\cal
A}^\mu({\bf p},{\bf k})f({\bf p}) \sqrt{P_0/p_0}(dp/dP)$. The
factor $\sqrt{P_0/p_0}\,dp/dP$ arises due to the change of the
integration variables from ${\bf p}$ to ${\bf P}$. One can regard
the function $F({\bf p})$ as the one-particle wave function in the
zero-photon sector in the ${\bf p}$-representation and the
function $G^\mu({\bf p},{\bf k})$ as that in the one-photon sector
with a photon with momentum $\hbar {\bf k}$ in the ${\bf
P}$-representation. This suggests that the expectation value of
the $z$-coordinate is
\begin{eqnarray}
\langle z\rangle & = & \frac{i\hbar}{2} \int\frac{d^3{\bf
p}}{(2\pi\hbar)^3} F^*({\bf p})
\stackrel{\leftrightarrow}{\partial}_p F({\bf p})\nonumber \\ &&
-\frac{i}{2} \int\frac{d^3{\bf k}}{2k_0(2\pi)^3} \int\frac{d^3{\bf
P}}{(2\pi\hbar)^3} G^{\mu *}({\bf p},{\bf k})
\stackrel{\leftrightarrow}{\partial}_P G_\mu({\bf p},{\bf k})\,.
\nonumber \\ \label{FG}
\end{eqnarray}
We will present a justification of this formula
elsewhere~\cite{HM2}. Since $f({\bf p})$ is
real, we have
\begin{eqnarray}
\langle z\rangle & = & -\hbar \int\frac{d^3{\bf p}}{(2\pi\hbar)^3}
| f({\bf p})|^2
\partial_p {\rm Re}\,{\cal F}({\bf p})\nonumber \\
&&  -\frac{i}{2} \int \frac{d^3{\bf p}}{(2\pi\hbar)^3} | f({\bf
p})|^2 \nonumber \\ && \times \int \frac{d^3{\bf k}}{2k_0(2\pi)^3}
{\cal A}^{\mu *}({\bf p},{\bf k})
\stackrel{\leftrightarrow}{\partial}_p {\cal A}_\mu({\bf p},{\bf
k})\,.
\end{eqnarray}
We have let ${\bf P} \to {\bf p}$ in
the second term because this term
will turn out to be of order $\hbar^0$.
(Note that $(P_0,{\bf P}) = (p_0,{\bf p})$ at this order because
the photon energy $\hbar k_0$ and momentum $\hbar {\bf k}$ are of
order $\hbar$.) Then, using the
assumption that the function $f({\bf p})$ is sharply peaked, we
find for small $\hbar$
\begin{eqnarray}
\langle z\rangle & = & - \hbar\, \partial_p{\rm Re}\,{\cal F}({\bf
p})\nonumber \\ && - \frac{i}{2} \int \frac{d^3{\bf
k}}{2k_0(2\pi)^3} {\cal A}^{\mu *}({\bf p},{\bf k})
\stackrel{\leftrightarrow}{\partial}_p {\cal A}_\mu({\bf p},{\bf
k})\,,
\end{eqnarray}
where ${\bf p} = (0,0,p)$ is now the momentum where the function
$f$ has the peak. The first term is the contribution from the
one-loop correction to the potential rather than from radiation
reaction. Thus, the second term is identified as the position
shift due to radiation reaction.  Therefore the position shift in
quantum field theory is
\begin{equation}
\delta z_{\rm Q} = -\frac{i}{2} \int \frac{d^3{\bf
k}}{2k_0(2\pi)^3} {\cal A}^{\mu *}({\bf p},{\bf k})
\stackrel{\leftrightarrow}{\partial}_p {\cal A}_\mu({\bf p},{\bf
k})\,. \label{shift}
\end{equation}
The photon-emission amplitude ${\cal A}^\mu({\bf p},{\bf k})$ can
be obtained to leading order in $\hbar$ by using the WKB wave
function (\ref{WKB}) as in Ref.~\cite{Higuchi2}.  The result turns
out to be identical with the amplitude for a classical charged
particle with final momentum ${\bf p}=(0,0,p)$ which passes
through the spacetime point $(t,x,y,z)=(0,0,0,0)$.  If the 
coordinates of
this particle are $X^\mu(\tau)$ at proper time $\tau$, then the
classical current generated by this charge is $j^\mu(t,{\bf
x})=e(dX^\mu/dt)\delta[{\bf x}-{\bf X}(\tau)]$. Thus, we find
\begin{eqnarray}
{\cal A}^\mu({\bf p},{\bf k}) & = & - \int d^4 x e^{ik_0 t - i{\bf
k}\cdot {\bf x}} j^\mu(t,{\bf x})\nonumber \\ & = & -
e\int_{-\infty}^{+\infty} d\tau \frac{dX^\mu}{d\tau} e^{ik^0 X^0 -
i{\bf k}\cdot {\bf X}}\,. \label{first}
\end{eqnarray}
Note that the right-hand side depends on $p$ through the world
line $X^\mu(\tau)$.  We write $X^\mu = (t,0,0,z)$ from now on. Let
$\theta$ be the angle between the $z$-direction and the vector
${\bf k}$ and define $\xi \equiv t -  \hat{n}\cdot {\bf x} = t -
z\cos\theta$, where $\hat{n} \equiv {\bf k}/k_0$.  Then 
Eq.~(\ref{first}) can be written
\begin{eqnarray}
{\cal A}^\mu({\bf p},{\bf k}) & = &  - e\int_{-\infty}^{+\infty}
d\xi\, \frac{dX^\mu}{d\xi} e^{ik\xi}\chi(\xi)\,, \label{ill}\\ & =
& \frac{e}{ik}\int_{-\infty}^{+\infty} d\xi\,\frac{d\
}{d\xi}\left[\frac{dX^\mu}{d\xi}\chi(\xi)\right] \,e^{ik\xi}\,,
\label{ill2}
\end{eqnarray}
where $k \equiv k_0$. We have inserted a function $\chi(\xi)$
which takes the value one in the region with $d^2 X^\mu/d\xi^2
\neq 0$ and decreases to zero smoothly for large $|\xi|$ to
regularize the integral. By using Eq.~(\ref{ill2}) for ${\cal
A}^\mu({\bf p},{\bf k})$ and Eq.~(\ref{ill}) for $\partial_p
{\cal A}^\mu({\bf p},{\bf k})$ in Eq.~(\ref{shift}) we
find~\cite{HM2}
\begin{equation}\label{quantshifteqn}
  \delta z_{\rm Q} = -\frac{\alpha_c}{4\pi}\int d\Omega \int d\xi
\frac{d^2X^{\mu}}{d\xi^2}
  \frac{\partial}{\partial p} \left( \frac{dX_{\mu}}{d\xi} \right)\,,
\end{equation}
where $d\Omega$ is the integral over the solid angle for ${\bf
k}$. (The $p$-derivative is taken with $\xi$ fixed.) Thus, the
position shift $\delta z_{\rm Q}$ is independent of the function
$\chi(\xi)$.

In order to show that the quantum position shift
$\delta z_{\rm Q}$ agrees with its classical
counterpart, we first note by using $d\xi/dt = 1 -
\dot{z}\cos\theta$ that
\begin{eqnarray}\label{twoeval}
  \frac{d^2z}{d\xi^2} &=&
  \frac{\ddot{z}}{(1-\dot{z}\cos\theta)^3} \\
  \frac{d^2t}{d\xi^2} &=&
 \frac{\ddot{z}}{(1-\dot{z}\cos\theta)^3}\cos\theta
= \frac{d^2z}{d\xi^2}\cos\theta\,.
\end{eqnarray}
Next we obtain, by interchanging the order of differentiation,
\begin{equation}\label{pdiffxmu}
  \frac{\partial}{\partial p}\left(\frac{dX^{\mu}}{d\xi}\right) =
  \frac{dt}{d\xi}\frac{d}{dt}\left(\frac{\partial
  X^{\mu}}{\partial p}\right)_\xi\,.
\end{equation}
Then by the chain rule we find
\begin{equation}\label{chaintp}
  \left(\frac{\partial t}{\partial p}\right)_{\xi}
= \left(\frac{\partial z}{\partial p}\right)_{\xi}
  \left(\frac{\partial t}{\partial z}\right)_{\xi}
= \left(\frac{\partial z}{\partial p}\right)_{\xi}\cos\theta\,.
\end{equation}
These equations can be used to find
\begin{equation}
\frac{d^2X^{\mu}}{d\xi^2}
  \frac{\partial}{\partial p} \left( \frac{dX_{\mu}}{d\xi} \right)
= - \frac{\ddot{z}}{(1-\dot{z}\cos\theta)^4}
\frac{d}{dt}\left(\frac{\partial z}{\partial p}\right)_\xi
  \sin^2\theta\,.
\end{equation}
By substituting this expression in the quantum position shift
(\ref{quantshifteqn}) and changing the integration variable from
$\xi$ to $t$, we find
\begin{eqnarray}\label{3}
 \delta z_{\rm Q}
& = & -\frac{\alpha_c}{4\pi} \int d\Omega \int_{-\infty}^0 dt
\frac{d\ }{dt}\left[
\frac{\ddot{z}}{(1-\dot{z}\cos\theta)^3}\right] \nonumber \\ && \
\ \ \ \ \ \ \ \ \ \ \ \ \ \ \ \ \ \ \ \ \ \ \ \ \ \ \ \ \ \times
\left(\frac{\partial z}{\partial p}\right)_\xi
  \sin^2\theta\,,
\end{eqnarray}
where we have changed the integration range from
$(-\infty,+\infty)$ to $(-\infty, 0]$ because $\ddot{z}=0$ for
$(0,+\infty)$, and then integrated by parts.

By substituting $dt = d\xi + \cos\theta\,dz$ in the relation $dz =
\dot{z}\,dt + (\partial z/\partial p)_t\, dp$ and solving for $dz$, we
obtain
\begin{equation}
\left(\frac{\partial z}{\partial p}\right)_\xi =
\frac{1}{1-\dot{z}\cos\theta}
\left(\frac{\partial z}{\partial p}\right)_t\,.
\end{equation}
By using this expression in Eq.~(\ref{3}) and performing the
$\Omega$-integration, we obtain
\begin{eqnarray}
\delta z_{\rm Q} &=& -\frac{2\alpha_c}{3} \int_{-\infty}^0 dt\,
\left[
\gamma^4\frac{d\ddot{z}}{dt}
+3\gamma^6\ddot{z}^2\dot{z}\right]\left(\frac{\partial
z}{\partial p}\right)_t \nonumber
\\ &=& - \int_{-\infty}^0 dt\, F_{\rm LD} \left(\frac{\partial
z}{\partial p}\right)_t\,, \label{posshift}
\end{eqnarray}
where $F_{\rm LD}$ is defined by Eq.~(\ref{LDLD}).

In order to interpret this formula let us consider the following
equation:
\begin{equation}\label{general}
  m\frac{d^2z}{d\tau^2}=\left[ F_{\rm ext} (z,t) + F_{\rm LD} \right]
\frac{dt}{d\tau}\,.
\end{equation}
Our model is a special case with $F_{\rm ext}(z,t)=-V'(z)$. Define
$P \equiv m\,dz/d\tau$ and let $(z,P)=(z_0(t),P_0(t))$ be a
solution of this equation with $F_{\rm LD}$ set to zero and
$(z,P)=(z_0+\Delta z,P_0+\Delta P)$ be a linearized solution about
$(z_0,P_0)$ with $F_{\rm LD}$ set to zero. Then $\Delta z$ and
$\Delta P$ satisfy
\begin{eqnarray}\label{homogenshifts1}
\frac{d}{dt}\Delta z &=& m^{-1}(1-\dot{z}^2_0)^{3/2}\Delta P
\equiv A(t) \Delta P\,,
\\ \frac{d}{dt}\Delta P &=& \frac{\partial F_{\rm ext}}{\partial
z}\Big|_{z=z_0} \Delta z \equiv B(t) \Delta z\,.
\label{homogenshifts2}
\end{eqnarray}
Let $(\Delta z_s(t),\Delta P_s(t))$ be a solution of these
equations satisfying $(\Delta z_s(s),\Delta P_s(s))=(0,1)$. Then
$(\partial z/\partial p)_t = \Delta z_0(t)$.  Hence 
Eq.~(\ref{posshift}) becomes
\begin{equation}
\delta z_{\rm Q} = - \int_{-\infty}^0 dt\,F_{LD}(t)\Delta
z_0(t)\,.
\end{equation}
Now, the approximate solution $(z_0+\delta z, P_0 + \delta P)$ of
Eq.~(\ref{general}) with $F_{\rm LD}\neq 0$ obtained by
perturbing $(z_0,P_0)$ to first order in $F_{\rm LD}$ satisfies
\begin{eqnarray}\label{inhomogenshifts1}
\frac{d}{dt}\delta z &=& A(t) \delta P\,,
\\
\frac{d}{dt}\delta P &=& B(t) \delta z + F_{\rm LD}(t)\,,
\label{inhomogenshifts2}
\end{eqnarray}
where $F_{\rm LD}(t) \equiv F_{\rm LD}|_{z=z_0}$.  The solution of
these equations with the initial condition $\lim_{t\to
-\infty}(\delta z,\delta P) = (0,0)$ is given by
\begin{eqnarray}
\delta z(t) & = & \int_{-\infty}^t ds\, F_{\rm LD}(s) \Delta
z_{s}(t)\,,\\ \delta P(t) & = & \int_{-\infty}^t ds\, F_{\rm
LD}(s) \Delta P_{s}(t)\,.
\end{eqnarray}
Hence the classical position shift is
\begin{equation}
\delta z = \int_{-\infty}^0 dt\,F_{\rm LD}(t)\Delta z_{t}(0)\,.
\end{equation}
Thus, to establish that the position shift $\delta z_{\rm Q}$ in
QED at order $e^2$ in the limit $\hbar \to 0$ 
is equal to that in the Lorentz-Dirac theory,
$\delta z$, we only need to establish $\Delta z_0(t) = - \Delta
z_t(0)$.  In fact one can show in general that $\Delta
z_s(t)=-\Delta z_t(s)$ as follows. Note first that due to
Eqs.~(\ref{homogenshifts1}) and (\ref{homogenshifts2})
the symplectic
product defined by
\begin{equation}\label{8}
\langle \Delta z^{(1)},\Delta P^{(1)}|\Delta z^{(2)},\Delta
P^{(2)} \rangle \equiv \Delta z^{(1)}\Delta P^{(2)} - \Delta
P^{(1)}\Delta z^{(2)}
\end{equation}
is time-independent for any pair of solutions $(\Delta
z^{(1)},\Delta P^{(1)})$ and $(\Delta z^{(2)},\Delta P^{(2)})$.
Hence we have
\begin{eqnarray}
&& \Delta z_s(t)\Delta P_t(t) - \Delta P_s(t) \Delta z_t(t)
\nonumber \\
&&\ \ \ \ \ 
=  \Delta z_s(s)\Delta P_t(s)
  - \Delta P_s(s) \Delta z_t(s)\,.
\end{eqnarray}
Since $\Delta z_s(s) = \Delta z_t(t) = 0$ and $\Delta P_s(s)=
\Delta P_t(t)=1$, we have $\Delta z_s(t) = -\Delta z_t(s)$ as
required. Hence $\delta z_{\rm Q} = \delta z$.

We have shown the equality of the classical and quantum position
shifts at order $e^2$ in the limit $\hbar \to 0$
for any system described by
Eq.~(\ref{general}) {\em provided that the photon emission amplitude
is given by Eq.~(\ref{first})}.  We have derived 
Eq.~(\ref{first}) only for the case where $F_{\rm ext}(z,t)$ is
$t$-independent although
it appears likely to hold in a more general case.  Let us verify
that the classical position shift $\delta z$ with 
$F_{\rm ext}(z,t)=-V'(z)$ given by 
Eq.~(\ref{deltaz}) is indeed equal to $\delta z_{\rm Q}$ in 
Eq.~(\ref{posshift}), though our general argument guarantees the
agreement. To this end we need to calculate $(\partial z/\partial
p)_t$. Since the energy is conserved, we have
\begin{equation}\label{11}
  \sqrt{m^2\left(dz/d\tau\right)^2 + m^2}+V(z) =
  \sqrt{p^2+m^2}\,.
\end{equation}
Thus
\begin{equation}
\frac{dz}{dt} =
\left[1-m^2\left(\sqrt{p^2+m^2}-V\right)^{-2}\right]^{1/2}\,.
\end{equation}
By integrating this with the condition $z=0$ at $t=0$ and
differentiating with respect to $p$, we find
\begin{equation}
\left(\frac{\partial z}{\partial p}\right)_t =
\frac{\dot{z}|_0}{m}\int^t_{0}\frac{1}{\gamma^3(t')[\dot{z}(t')]^2}
dt'\,,  \label{ooo}
\end{equation}
where $\dot{z}|_0 = p/\sqrt{p^2+m^2}$ is the final velocity.
The substitution of Eq.~(\ref{ooo}) in Eq.~(\ref{posshift}) gives the
quantum position shift as follows:
\begin{equation}
\delta z_{\rm Q} = -\frac{\dot{z}|_0}{m}\int^0_{-\infty}
\left(\int^{t}_{0}\frac{1}{\gamma^3(t')[\dot{z}(t')]^2}dt'
\right)F_{\rm LD}\frac{dz}{dt}dt\,,
\end{equation}
which is identical with the position shift $\delta z$ in the
Lorentz-Dirac theory given by Eq.~(\ref{deltaz}).

In this Letter we have reported that the $\hbar \to 0$ limit of 
the position shift due to
radiation reaction in QED to lowest non-trivial order in $e^2$ 
is correctly reproduced by the Lorentz-Dirac theory
for linear acceleration.  It would be interesting to extend our
result to three dimensional motion.  
Investigation of the one-loop correction to the potential 
is also important since this also
affects the motion of the particle.  The corresponding
correction to the position of the particle in a model with
space-dependent mass term~\cite{Higuchi2} was found to be of order
$\hbar^{-1}$, overwhelming the effect from radiation
reaction.  However, in the more realistic model we have considered 
here, the one-loop correction to the position appears to be
at most of order $\hbar^0$.  It
would be very interesting to estimate this contribution and
compare it with that from radiation reaction.

\end{document}